\documentclass[9pt,conference]{IEEEtran}

\usepackage[preprint]{waspaa25}

\usepackage{bm} 

\usepackage{xcolor}
\usepackage{multirow}

\providecommand{\nz}[1]{\textcolor{black}{{#1}}} 

\title{
Towards Perception-Informed Latent HRTF Representations
}

\name{
    You Zhang$^{1,2}$\sthanks{Work partially performed while You Zhang was a research intern at Meta}, 
    Andrew Francl$^{2}$, 
    Ruohan Gao$^{3}$, 
    Paul Calamia$^{2}$, 
    Zhiyao Duan$^{1}$, 
    Ishwarya Ananthabhotla$^{2}$
}
\address{$^{1}$University of Rochester \;  $^{2}$Meta Reality Labs Research \; $^{3}$University of Maryland, College Park
}

\begin{document}

\maketitle

\begin{abstract}

Personalized head-related transfer functions (HRTFs) are essential for ensuring a realistic auditory experience over headphones, because they take into account individual anatomical differences that affect listening.  Most machine learning approaches to HRTF personalization rely on a learned low-dimensional latent space to generate or select custom HRTFs for a listener.  However, these latent representations are typically learned in a manner that optimizes for spectral reconstruction but not for perceptual compatibility, meaning they may not necessarily align with perceptual distance.  In this work, we first study whether traditionally learned HRTF representations are well correlated with perceptual relations using auditory-based objective perceptual metrics; we then propose a method for explicitly embedding HRTFs into a perception-informed latent space, leveraging a metric-based loss function and supervision via Metric Multidimensional Scaling (MMDS).  Finally, we demonstrate the applicability of these learned representations to the task of HRTF personalization. We suggest that our method has the potential to render personalized spatial audio, leading to an improved listening experience.


\end{abstract}

\section{Introduction}
\label{sec:intro}

Personalized head-related transfer functions (HRTFs) are crucial for spatial audio rendering, enabling accurate 3D sound reproduction over headphones for any listener~\cite{Xie14, herre2015MPEG}.  Since HRTFs differ across individuals due to varying anatomical features, personalizing them is essential to achieving realistic spatial audio experiences. However, measuring HRTFs for each listener is time-consuming, resource-intensive, and often impractical outside of specialized lab environments~\cite{li2020measurement}.  Learning-based approaches that attempt to predict HRTFs from sparse measurements or human geometry have recently emerged, but it has remained a long-standing research problem.  Learning-based personalization is challenging in part due to the high intrinsic dimensionality associated with HRTFs, as frequency responses must be modeled over hundreds of spatial locations for accurate sound rendering.  To this end, learning efficient latent representations from measured HRTFs is a key requirement.  

Recent deep learning research has pushed HRTF representation learning towards practical personalization~\cite{fantini2025survey}.  These representations can be used to reconstruct complete HRTFs, perform spatial interpolation, and predict HRTFs for listeners based on their anthropometric features. \textit{Autoencoder-based approaches} first compress the complete HRTFs into a low-dimensional latent code, which is then linked to anthropometric inputs or used for spatial up‑sampling~\cite{yamamoto2017fully, chen2019autoencoding, ito2022head, zurale2023learning, zhao2025head}. \nz{These models are usually tied to the spatial sampling grid of the training database and can only reconstruct or upsample to predetermined directions unless explicit source position conditioning is applied~\cite{ito2022head}}.  
\textit{Implicit neural representations} (INRs) address this limitation by treating the HRTF as a continuous function: a neural network learns the mapping from a spatial coordinate to the corresponding magnitude~\cite{gebru2021implicit, zhang2023hrtf}.  Thanks to their grid-agnostic nature~\cite{sitzmann2020implicit}, INRs enable automatic interpolation across databases with different spatial sampling schemes. Recent work further exploits INRs for high-resolution upsampling, personalized HRTF prediction, and binaural cue generation~\cite{di2024neural, lobato2024process, masuyama2025retrieval, lu2025bicg}.

Despite recent progress, it remains uncertain whether existing latent representations truly align with human auditory perception. The gold standard way to assess perceptual validity is through subjective listening tests~\cite{yamamoto2017fully, zhang2020spatialPCA, zhao2025head}, which are time-consuming and impractical for rapid model iteration. In practice, most representation learning methods only optimize a spectral distance loss and report the same as a metric for evaluation. Although this produces a close spectral match, it does not guarantee perceptual similarity: models often smooth the HRTF magnitude across directions, especially on high-frequency bins~\cite{zhang2020spatialPCA, wang2021global}, leaving residual spectral errors that can obscure audible cues. Consequently, the HRTF that yields the smallest spectral distance is not always the one perceived as most similar in listening tests~\cite{doma2023examining}.  Hence, we aim to bridge this gap by developing methods that optimize perceptual alignment \emph{in addition to} spectral reconstruction, enabling more efficient—and perceptually accurate—HRTF personalization.



Several recent efforts~\cite{andreopoulou2018comparing, pelzer2020head, ananthabhotla2021framework, Marggraf2024hrtf, lobato2024process, yao2024perceptually} have begun to incorporate some perceptual facets into HRTF modeling—for example, a deep-learning metric modeling localization error~\cite{ananthabhotla2021framework}, an HRTF recommendation engine that predicts coloration~\cite{Marggraf2024hrtf}, and a perceptually enhanced spectral distance metric~\cite{yao2024perceptually}. More broadly, in perceptual studies of personalized HRTFs, the 
metrics for evaluating preference over a generic HRTF are commonly grouped into three dimensions: \emph{coloration}, \emph{externalization}, and \emph{localization}~\cite{majdak2013sound, brinkmann2017authenticity, jenny2020usability}. However, it is still unclear whether these perceptual dimensions are captured in existing representation learning methods, and how to further align the learned latent representations with perception beyond these baselines.  

\textbf{Our contribution.}  
We systematically study whether latent HRTF representations preserve perceptual relations by embedding HRTFs into a metric space informed by objective computational metrics. Specifically, we (i) compare standard representation learning pipelines on perceptual relation preserving,
and (ii) introduce Metric Multidimensional Scaling (MMDS) supervision and perceptual loss terms to align the latent space with these perceptual metrics. (iii) We show that the resulting representations exhibit significantly higher correlations with localization, coloration, and externalization measures, and we demonstrate their practical utility for HRTF personalization.  





\section{Do Learned Latent HRTF Representations Preserve Perceptual Relations?}

\subsection{Preliminary: Computational Auditory Modeling}
\label{ssec:perceptualMetrics}

Computational auditory models~\cite{meddis2010computational} translate raw acoustic signals into feature spaces that approximate the cues the human auditory system processes—such as interaural level/phase differences, spectral notches. When these feature transformations are combined with decision rules tuned and validated against listening experiments, they yield \emph{objective perceptual metrics}: algorithmic predictions that track listener judgments with high reliability.
In spatial audio research, such metrics have been developed to model the perception of sound coloration~\cite{McKenzie2022pbc}, externalization~\cite{baumgartner2021decision}, and localization~\cite{francl2022deep, barumerli2023bayesian}. These metrics are grounded in psychoacoustic evidence and are pre-computable. 


\textbf{Predicted Binaural Coloration (PBC)}~\cite{McKenzie2022pbc} is developed to calculate the perceived coloration difference between two sets of signals. We adapt this model for HRTF coloration to input two HRTF magnitude spectra. 
The HRTF magnitudes are weighted according to inverse equal loudness contours, 
then converted to the sone scale~\cite{zwicker2013psychoacoustics}, 
and further weighted according to its equivalent rectangular bandwidth.

\textbf{Auditory Externalization Perception (AEP)}~\cite{baumgartner2021decision} calculates perceived externalization by modeling how spectral cues contribute to sound perception in static listening environments. The findings in~\cite{baumgartner2021decision} suggest that the best predictions were made using a fixed ratio of 60\% monaural spectral similarity and 40\% interaural spectral similarity of interaural level differences. 
We convert this similarity to distance by subtracting it from 1, since self-similarity is 1 (zero distance).

\textbf{Difference of Root Mean Square Error in Polar Angles (DRMSP)} was one of several localization metrics reported in the Listener Acoustic Personalization (LAP) Challenge Task 1 validation~\cite{geronazzo2024technical}\nz{, where a threshold of 5.90 degrees was applied. Derived from an auditory localization model~\cite{barumerli2023bayesian}, the metric quantifies the increase in root mean square errors in polar angles when localizing with a non-individual HRTF compared to an individual one~\cite{middlebrooks1999virtual}}. As our representation learning models use only the magnitude of the HRTF (discarding phase), they do not encode cues that dominate lateral localization. Hence, we adopt this DRMSP metric as a magnitude-sensitive measure of the distance in localization perception.

\subsection{Experimental Setup}
\label{ssec:case_exp_setup}

\textbf{Dataset}: Sound Sphere 2 (SS2)~\cite{warneckeHRTF2024} is a high-resolution HRTF database consisting of 78 subjects and $L=$1,625 measurement locations. The HRIRs were recorded at a high sampling rate of 48 kHz in a purpose-built anechoic chamber with a motorized arc setup, ensuring precise measurement of sound directions. We divided the subjects into 65 subjects for training and 13 subjects for evaluation.

\textbf{Models}: We selected two representative HRTF representation learning architectures to train for reconstructing HRTF magnitudes across all locations. The HRTF data are represented as $\mathbf{x} \in \mathbb{R}^{L \times K \times 2}$, where $K$ is the number of frequency bins. A convolutional autoencoder (CAE), adapted from~\cite{zhao2025head}, transforms the HRTF into $\mathbf{z} \in \mathbb{R}^{D \times K \times 2}$. The model includes an encoder with two layers of 2D convolutional neural networks, followed by another two convolutional layers as the decoder to reconstruct the HRTF magnitudes.
We adapt the implicit neural representation (INR) model~\cite{zhang2023hrtf} to model HRTFs for both ears as $G(\theta, \phi, \mathbf{z})$, where the sound direction of azimuth and elevation angles $(\theta, \phi)$ and the latent code $\mathbf{z} \in \mathbb{R}^D$ are mapped to an individual's HRTFs as $G: \mathbb{R}^{2 + D} \mapsto \mathbb{R}^{K \times 2}$. The model consists of two layers of multilayer perceptrons (MLPs), each with 2048 hidden units.

\textbf{Implementation details}: The linear-domain HRTFs are obtained by taking a 256-point fast Fourier transform (FFT) of the HRIRs following~\cite{chen2019autoencoding, zhang2023hrtf}, and we filter the HRTFs to retain frequencies between 200 Hz and 16 kHz. This results in $K = 85$ frequency bins for each ear. We implemented the PBC~\cite{McKenzie2022pbc}, AEP~\cite{baumgartner2021decision} and localization~\cite{barumerli2023bayesian} models from the MATLAB Auditory Modeling Toolbox (AMT)~\cite{majdak2022amt}. The DRMSP was implemented from the LAP challenge\footnote{\url{https://github.com/Audio-Experience-Design/LAPChallenge/blob/main/task1/task1_validate.m}}.
For the representation learning models, we train the models for 300 epochs using $L_2$ reconstruction loss only and choose the epoch with the lowest reconstruction loss on the test set for evaluation.

\textbf{Evaluation}:
To evaluate the extent to which the learned representations preserve perceptual relation, we use the Pearson correlation between the latent space distance of paired subjects and the perceptual metrics between the two subjects' HRTFs. 
\begin{equation}
    \rho_{A,B} = \frac{\mathbb{E}[(A - \mu_A)(B - \mu_B)]}{\sigma_A \sigma_B},
\end{equation}
where $A$ is the set of latent space distances and $B$ is the set of perceptual metric values (e.g., PBC) between the HRTFs. For subjects $i, j \in \{1, 2, \dots, N\}$, we compute $L_2$ distance in the latent space as $\| \mathbf{z}_i - \mathbf{z}_j \|_2$, as $L_2$ distance is effective for capturing the geometric structure of data that is normally distributed or approximately so. The distance in the perceptual metrics space is $m_{i,j} = f(\mathbf{x}_i, \mathbf{x}_j)$, where $f$ is any of the computational auditory models introduced in Section~\ref{ssec:perceptualMetrics}.

We use Spectral Difference Error (SDE) to evaluate the spectral distance~\cite{warnecke2022hrtf}. The SDE for each frequency bin with index $k$ is:
\begin{equation}
\mathrm{SDE}_k(\operatorname{H}, \hat{\operatorname{H}})=\frac{1}{L} \sum_{\theta, \phi} \left| 20 \cdot \log _{10} \left( \frac{\operatorname{H}(\theta, \phi, k)}{\hat{\operatorname{H}}(\theta, \phi, k)} \right) \right|,
\label{eq: sde}
\end{equation}
where $\operatorname{H}(\theta, \phi, k)$ and $\hat{\operatorname{H}}(\theta, \phi, k)$ indicate the linear-scale magnitude of the ground-truth HRTF and the reconstructed HRTF of the $k$-th frequency bin at the direction $(\theta, \phi)$, respectively. The median SDE across all frequency bins was computed to obtain a single SDE value. 


We computed the pairwise distances for the three perceptual metrics (PBC, AEP, DRMSP) and the spectral distance (SDE) between the ground-truth HRTFs in SS2. The Pearson correlations were 0.78, 0.73, and 0.37, respectively, indicating that SDE has a strong correlation with PBC and AEP but only a weak correlation with DRMSP.

\begin{table}[]
\centering
\caption{Pearson correlation results for three perceptual metrics. 
}
\label{tab:case}
\sisetup{
    reset-text-series = false, 
    text-series-to-math = true, 
    mode=text,
    tight-spacing=true,
    round-mode=places,
    round-precision=2,
    table-format=2.2,
    table-number-alignment=center
}
\begin{tabular}{cc|ccc}
\toprule
         Models            &   Partitions    & \textbf{PBC} $\downarrow$       & \textbf{AEP} $\downarrow$      & \textbf{DRMSP} $\downarrow$     \\ \midrule 
\multirow{2}{*}{CAE} & train & 0.60±0.11  & 0.71±0.08 & 0.43±0.13  \\
                     & test  & -0.15±0.21 & 0.07±0.31 & -0.10±0.27 \\ \midrule
\multirow{2}{*}{INR}  & train & 0.60±0.09  & 0.60±0.14 & 0.40±0.15  \\
                     & test  & 0.71±0.22  & 0.55±0.23 & 0.41±0.27    \\
  \bottomrule                   
\end{tabular}
\end{table}

\begin{figure}[t]
  \centering
  \centerline{\includegraphics[width=1.01\columnwidth]{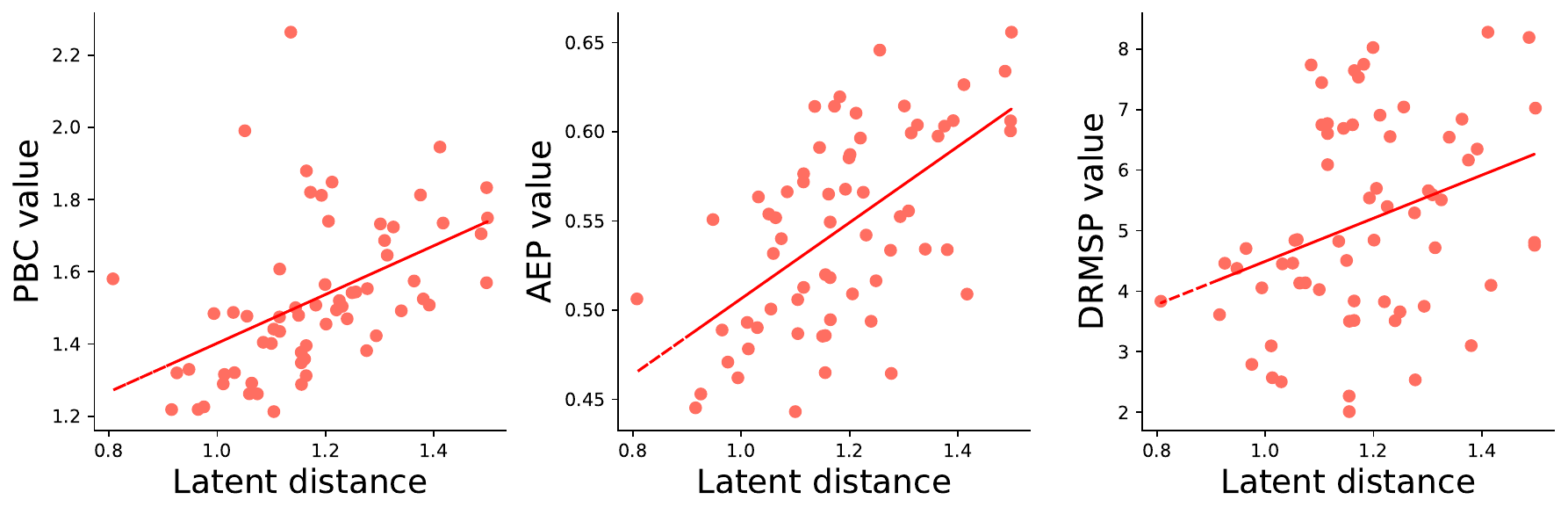}}
  \caption{Correlation between latent space and the perceptual metrics (PBC, AEP, DRMSP from left to right) of the INR model on the training subjects with Subject \texttt{IXJ675081622} serving as the anchor. 
  }
  \label{fig:pearson}
\end{figure}

\subsection{Experimental Results and Analyses}


We set an anchored subject for each trial and compute the $L_2$ distance between the latent $\mathbf{z}$ of the anchored subject and each of the other subjects, and we have calculated the pairwise perceptual metrics between the ground-truth (GT) HRTFs of the subjects.






\begin{table*}[]
\centering
\caption{Comparison of Pearson correlation of latent distances with the PBC metric and reconstruction error for the proposed methods and the baseline. 
}
\label{tab:main}
\begin{tabular}{llcccccccc}
\toprule
      &   \multirow{3}{*}{\textbf{Methods}}  & \multicolumn{4}{c}{\textbf{Pearson Correlation} $\uparrow$}                    & \multicolumn{4}{c}{\textbf{Reconstruction Error}}               \\
      &     & \multicolumn{2}{c}{Ground-truth (GT)} & \multicolumn{2}{c}{Reconstructed} & \multicolumn{2}{c}{SDE (dB) $\downarrow$} & \multicolumn{2}{c}{PBC $\downarrow$} \\ \cmidrule(lr){3-4}\cmidrule(lr){5-6} \cmidrule(lr){7-8}\cmidrule(lr){9-10}
      &     & train      & test      & train           & test            & train          & test        & train       & test      \\ \midrule
\textbf{Proposed}                    & $L_2+L_\text{Align}+L_\text{PBC}$ & \textbf{0.93±0.02}  & \textbf{0.80±0.14} & \textbf{0.95±0.01}       & \textbf{0.86±0.13}       & 0.87           & 1.58        & \textbf{0.56}        & 1.04      \\
Baseline                        & $L_2$         & 0.60±0.09  & 0.71±0.22 & 0.78±0.06       & 0.80±0.14       & \textbf{0.82}           & \textbf{1.51}        & 0.67        & 1.09    \\ \hline
\multirow{2}{*}{Ablation study} & $L_2+L_\text{Align}$     & 0.96±0.01  & 0.78±0.14 & 0.87±0.04       & 0.82±0.13       & 1.00           & 1.58       & 0.79        & 1.11      \\
& $L_2+L_\text{PBC}$     & 0.64±0.10  & 0.71±0.21 & 0.77±0.08       & 0.83±0.17       & 1.03           & 1.58        & 0.64        & \textbf{1.02}      \\
\bottomrule
\end{tabular}
\end{table*}

\begin{figure}[t]
  \centering
  \centerline{\includegraphics[width=0.99\columnwidth]{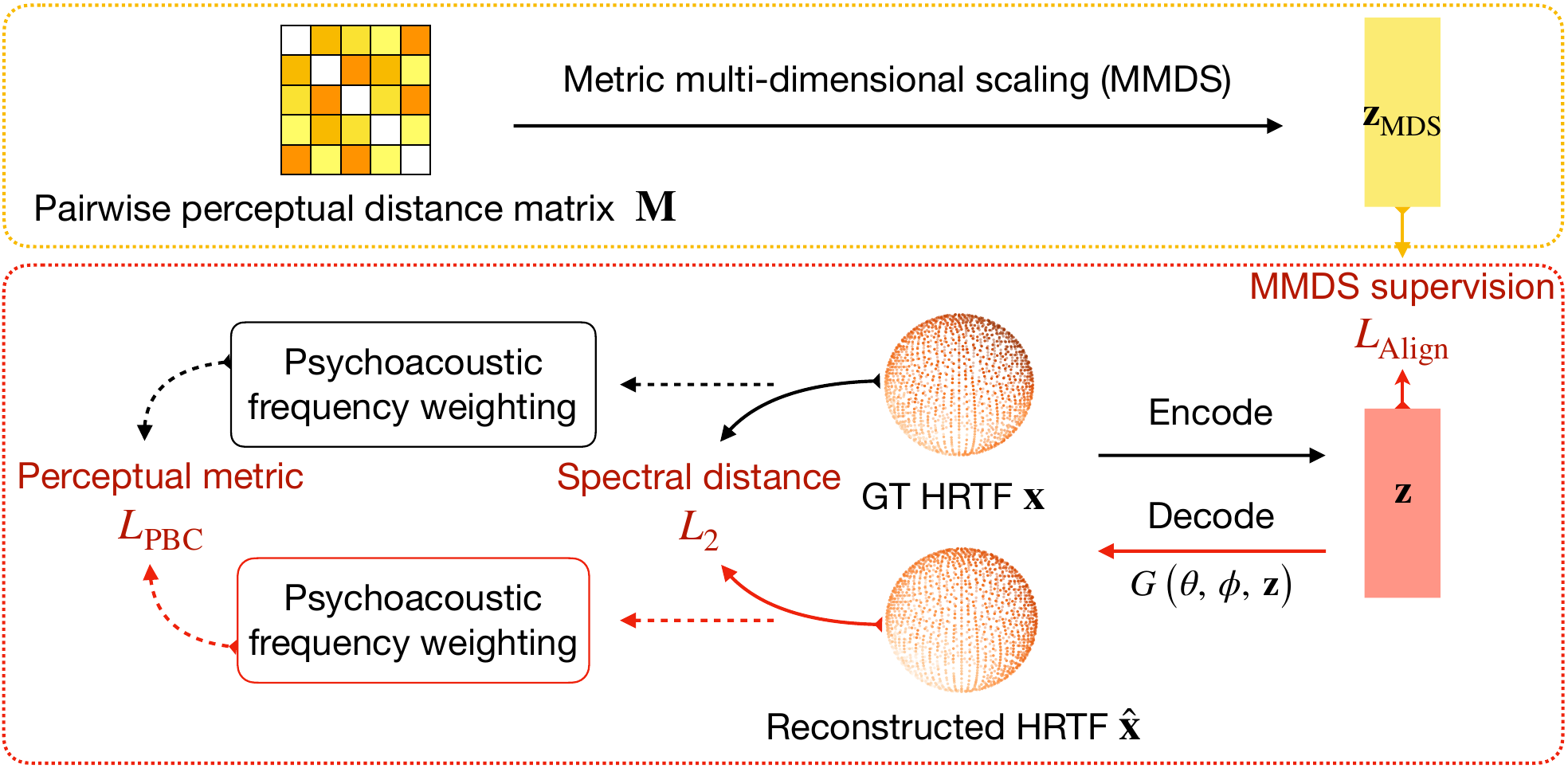}}
  \caption{Our pipeline of learning perception-informed HRTF representations.
  }
  \label{fig:pipeline}
\end{figure}

Table~\ref{tab:case} presents the Pearson correlation evaluation results for two models, CAE and INR, based on the three perceptual metrics introduced in Section~\ref{ssec:perceptualMetrics}. The respective standard deviations are across different anchored subjects in the training set for both the training and test configurations. On the training set, both models perform similarly, with values closely matching their Pearson correlation with the SDE, as the training objective focuses on minimizing spectral reconstruction error. However, on the test set, the CAE model fails to generalize the correlation, while the INR model is able to generalize much better, possibly due to its generative and grid-agnostic nature, which enables it to capture the global characteristics of HRTF magnitudes.
Among the three metrics, the correlations for PBC and DRMSP are even higher than those on the training data, though not statistically significant, likely due to the model overfitting the reconstruction objective. 

Regarding the correlation values from the INR model on both sets, we observe a high correlation when the perceptual metric aligns with the spectral error SDE. This indicates that models trained to minimize spectral distance inherit perceptual correlation from the spectral domain. We expect these correlations could likely be improved by adjusting the objective to incorporate perceptual metrics, either as a prior or through a loss function, motivating our proposed approach.

\section{Aligning with Perception-Informed Space}

We design our methods using the PBC metric as an example and further evaluate its generalization ability in Section~\ref{ssec:generalize}. For representation learning, we use the INR model~\cite{zhang2023hrtf}, which originally employs implicit gradient-based optimization~\cite{bond2020gradient} to simultaneously learn the latent representation and the generative model. To incorporate constraints into the latent space during training, we adapt the training paradigm to generative latent optimization~\cite{bojanowski18glo}, which was also applied in~\cite{lobato2024process}. In this modified approach, the latent space for each training subject is randomly initialized and then optimized jointly with the model weights to find the best latent variables for each sample. During inference, we treat the optimization as an inverse problem, similar to GAN inversion~\cite{9792208}, where we optimize the latent vectors by minimizing the loss while keeping the model weights fixed.



Fig.~\ref{fig:pipeline} illustrates the overall pipeline of our proposed representation learning method, which includes two auxiliary losses in addition to the spectral reconstruction error: a latent distance loss derived from embeddings learned through multidimensional scaling, and an optional perceptual loss, applicable when the perceptual metric is differentiable.

\subsection{Metric Multidimensional Scaling (MMDS)}
\label{ssec:mmds}

A straightforward approach is to enforce a correlation between $L_2$ and perceptual distance in the learned latent space and penalize the lack of correlation. However, this requires integrating a perception metric directly into training, but auditory model-based metrics are often non-differentiable. To address this, we precompute the pairwise perceptual distance metrics as $\mathbf{M}_{N \times N}$ and use them as guidance for the representation learning, where $m_{i,j}$ representing the perceptual distance between the $i$-th and $j$-th subjects. 

Metric Multidimensional Scaling (MMDS)~\cite{saeed2018survey} is used to transform this distance matrix $\mathbf{M}_{N \times N}$ into a set of coordinates $\mathbf{z}_\text{MDS} \in \mathbb{R}^{D}$ for each of the $N$ subjects, where the Euclidean distances between the points in the new space approximate the original perceptual distances as closely as possible. MMDS accomplishes this by converting the distance matrix into a cross-product matrix, followed by eigen-decomposition, which is equivalent to principal component analysis. Unlike conventional MDS, which is primarily used for dimensionality reduction, MMDS does not fit the training data directly. Instead, it encodes the distance matrix into a new dimensional subspace. 

\subsection{Loss Functions}

The resulting representation $\mathbf{z}_\text{MDS}$ from Section~\ref{ssec:mmds} is used as a supervision of representation learning, ensuring that the learned representation is close to the embedded space of MMDS. This alignment ensures that the distance in the learned latent space preserves the pairwise relationships inherent in the perceptual metrics. 
\begin{equation}
    L_\text{Align} = \| \mathbf{z} - \mathbf{z}_\text{MDS} \|_2  
\end{equation}
When the perceptual metrics are differentiable, we apply a loss to encourage the perceptual distance between the reconstructed HRTF and the ground-truth HRTF. This only applies to PBC in this study.
\begin{equation}
L_{\text{PBC}} = \mathrm{PBC}(\mathbf{x}, \hat{\mathbf{x}})
\end{equation}
Total loss is a combination of reconstruction and perceptual alignment:
\begin{equation}
L = L_2 + \alpha L_\text{Align} + \beta L_{\text{PBC}}
\end{equation}


\section{Experiments on Improving Alignment}

\subsection{Experimental Setup}

The entire training process is implemented in PyTorch. We reimplemented PBC with SciPy~\cite{virtanen2020scipy} and PyTorch to enable its use as a loss function, as every processing step is differentiable.
We set hyperparameters with $\alpha = 0.3$, $\beta=0.2$. The other setup, dataset, and evaluation metrics are the same as in Section~\ref{ssec:case_exp_setup}.



\subsection{Objective Perceptual Correlation Evaluation on PBC}
\label{ssec:obj_exp}
Table~\ref{tab:main} shows the experimental results comparing our proposed methods on PBC with the baseline method. The results include Pearson correlation coefficients and reconstruction errors on both the training and test sets, for four training loss configurations of the INR model.

\textbf{Comparison with baseline on the training set}. When comparing our proposed method with the baseline, we observe a significant improvement in the Pearson correlation on the training GT and reconstructed HRTFs. This indicates that our proposed loss functions—specifically, the addition of PBC and MMDS supervision—help align the learned latent space with perceptual distances. Additionally, the PBC reconstruction error on the training data also decreases, highlighting the effectiveness of incorporating perceptual information into the model.
On the reconstructed HRTFs, we recalculate the pairwise perceptual metrics among them. The correlation between the reconstructed HRTFs and the latent representation improves and remains similar to the correlation with the ground truth. This suggests that the model is effectively reconstructing the HRTFs while maintaining a high perceptual correlation, even after reconstruction. 

\textbf{Generalize to subjects in the test set}.
When evaluating the test set (unseen subjects), the proposed model retains the advantage: correlation rises from 0.71 to 0.80 on ground truth and from 0.80 to 0.86 on reconstructions, with paired $t$-tests indicating significance ($p\ll0.01$). Interestingly, the correlation between the generated HRTFs from the model and the ground truth is often higher than that of the training data. This may be due to slightly higher reconstruction errors on the test set, but it also indicates that the generated HRTFs have a high perceptual correlation with the learned latent representations. Although SDE increases a bit, as expected when generalizing to new subjects, the lower PBC error and higher correlation than the baseline suggest that the perceptual structure and correlation learned in training transfer to unseen data. 

This result also suggests that, provided the latent vector $\mathbf{z}$ is accurately predicted (from sparse measurements or anthropometry), the synthesized HRTFs should retain perceptual similarity, a crucial and promising property for prediction‑based HRTF personalization.

\textbf{Ablation study}.
The bottom two rows of Table~\ref{tab:main} illustrate the effect of each proposed loss. Removing the PBC loss still results in a large correlation gain over the baseline, indicating that the MMDS supervision is the main driver of perceptual correlation. Conversely, enforcing only the PBC term shows modest improvement in Pearson correlation but achieves the lowest PBC reconstruction error. Combining both terms yields the highest correlations, suggesting PBC complements the alignment loss and refines perceptual consistency.





\subsection{Generalization to Other Metrics and Datasets}
\label{ssec:generalize}

\begin{table}[]
\centering
\caption{Generalization results with Pearson correlation and spectral reconstruction for the proposed methods on AEP and DRMSP perceptual metrics. 
}
\label{tab:generalize}
\sisetup{
    reset-text-series = false, 
    text-series-to-math = true, 
    mode=text,
    tight-spacing=true,
    round-mode=places,
    round-precision=2,
    table-format=2.2,
    table-number-alignment=center
}
\begin{tabular}{clcccc}
\toprule
                      &    \multirow{2}{*}{\textbf{Methods}}    & \multicolumn{2}{c}{\textbf{Pearson correlation}$\uparrow$} & \multicolumn{2}{c}{\textbf{SDE (dB)}$\downarrow$} \\ \cmidrule(lr){3-4}\cmidrule(lr){5-6}
                      &        & train GT             & test GT              & train       & test      \\ \midrule
\multirow{2}{*}{AEP}  & $L_2+L_\text{Align}$ & 0.76±0.09          & 0.67±0.16          & 1.09        & 1.65      \\
                      & $L_2$     & 0.60±0.14          & 0.55±0.23          & 0.82        & 1.51      \\ \midrule 
\multirow{2}{*}{DRMSP} & $L_2+L_\text{Align}$ & 0.96±0.02          & 0.70±0.20          & 0.91        & 1.74      \\
                      & $L_2$     & 0.40±0.15          & 0.41±0.27          & 0.82        & 1.51     \\
  \bottomrule                   
\end{tabular}
\end{table}

\begin{table}[]
\centering
\caption{Generalization results with Pearson correlation and spectral reconstruction for our proposed method on the HUTUBS dataset on the PBC metric. 
}
\label{tab:hutubs}
\begin{tabular}{lcccc}
\toprule
\multirow{2}{*}{\textbf{Methods}} & \multicolumn{2}{c}{\textbf{Pearson correlation}$\uparrow$} & \multicolumn{2}{c}{\textbf{SDE (dB)}$\downarrow$} \\ \cmidrule(lr){2-3}\cmidrule(lr){4-5}
                        & train GT       & test GT        & train       & test      \\ \midrule
$L_2+L_\text{Align}+L_\text{PBC}$              & 0.98±0.01      & 0.71±0.13      & 0.29        & 1.60      \\
$L_2$                      & 0.58±0.12      & 0.62±0.14      & 0.42        & 1.45     \\
  \bottomrule  
\end{tabular}
\end{table}



Table~\ref{tab:generalize} reports results for the AEP and DRMSP metrics on SS2 dataset, both of which are non-differentiable; hence we only apply MMDS supervision on top of the $L_2$ objective. In both cases, Pearson correlation rises on training and on unseen test subjects, while SDE increases slightly, mirroring the trade‑off observed in Table~\ref{tab:main}.
For AEP, the correlation values for the proposed method are lower than those for PBC and DRMSP. An inspection of the $\mathbf{z}_\text{MDS}$ learned from the AEP distance matrix reveals that it only yields 0.82 distance correlation, versus 0.98 for PBC and DRMSP, indicating that AEP is highly nonlinear and therefore MMDS itself provides weaker supervision.

Table~\ref{tab:hutubs} shows the results of applying the proposed method on PBC for the 440-location, 93-listener HUTUBS dataset~\cite{fabian2019hutubs}, split 78/15 for train/test. Pearson correlation with the PBC metric again highly improves with only a slight SDE increase on the test set but improves on the training set, consistent with the trends in Table~\ref{tab:main}. 

These findings confirm that our proposed latent perception-informed representations generalize across perceptual metrics and datasets.







\section{Application and Discussions}

\subsection{Personalized HRTF Selection}



\begin{table}[]
\centering
\caption{Comparison of the best and top 5 candidates for personalized HRTF selection based on the latent representations learned with our proposed perception-informed method or the baseline.}
\label{tab:personalization}
\resizebox{1.02\linewidth}{!}{
\begin{tabular}{clcc|cc}
\toprule
\multicolumn{1}{l}{\multirow{2}{*}{}} & \multirow{2}{*}{\textbf{Methods}} & \multicolumn{2}{c}{\textbf{Best candidate}} & \multicolumn{2}{c}{\textbf{Top 5 candidates}} \\ \cmidrule(lr){3-4}\cmidrule(lr){5-6}
\multicolumn{1}{l}{}                  &                          & Metrics$\downarrow$          & SDE (dB)$\downarrow$          & Metrics$\downarrow$           & SDE (dB)$\downarrow$          \\
\midrule
\multirow{2}{*}{PBC} & $L_2+L_\text{Align}+L_\text{PBC}$ & \textbf{1.21} & 2.11 & \textbf{1.31} & \textbf{2.17} \\
                     & $L_2$     & 1.30 & \textbf{2.07} & 1.38 & 2.19 \\
\hline \midrule
\multirow{2}{*}{AEP} & $L_2+L_\text{Align}$ & 0.49 & 2.17 & \textbf{0.50} & 2.27 \\
                     & $L_2$     & \textbf{0.48} & \textbf{2.07} & 0.51 & \textbf{2.19} \\
\hline \midrule
\multirow{2}{*}{DRMSP} & $L_2+L_\text{Align}$ & \textbf{3.20} & 2.12 & \textbf{3.61} & 2.26 \\
                     & $L_2$     & 4.21 & \textbf{2.07} & 4.42 & \textbf{2.19} \\
\bottomrule
\end{tabular}
}
\end{table}

We demonstrate the use of the learned perception-informed latent representation for HRTF personalization. While we do not directly perform personalization in this work, which involves additional encoder designs to generate $\mathbf{z}$ representations~\cite{fantini2025survey}, our approach shows its applicability. In Section~\ref{ssec:obj_exp}, we show that our method improves perceptual alignment in reconstructed HRTFs, which can enhance HRTF prediction from the latent space. This section focuses on selection-based methods to further demonstrate the approach.


Given a pool of candidate subjects and a well-trained encoder that maps any listener to a latent $\mathbf{z}$, we test whether our perception-informed representation yields better personalization than a baseline trained with $L_2$ loss only. For each of the 13 listeners in the SS2 test partition~\cite{warneckeHRTF2024}, we identify the nearest (and top‑5 nearest) out of 65 training subjects in latent space using either distance metric. We then assess both latent representations by comparing the perceptual metrics and SDE between the selected HRTFs and the GT HRTFs.

Table~\ref{tab:personalization} summarizes the personalization results. Adding the alignment and perceptual losses consistently lowers the enforced metric values for the selected HRTFs with slightly higher SDE. Improvements appear immediately for the best candidate in PBC and DRMSP, while AEP shows a modest advantage once the top 5 candidates are considered, possibly due to the greater nonlinearity of the AEP metric discussed also in Section~\ref{ssec:generalize}. While the absolute gains are modest and their perceptual salience remains to be confirmed, we expect these benefits are likely to grow with larger training datasets.










\subsection{Limitations and Future Work}

This study also has limitations that should be acknowledged. 1) Our current implementation predicts only the \emph{magnitude} of the HRTF and ignores phase information, thus excluding phase-dependent cues. \nz{2) The \emph{objective} perceptual metrics may not always align with \emph{subjective} perceptual experience. 3) The MMDS-based supervision assumes a \emph{symmetric} dissimilarity matrix, whereas real-world perceptual degradation may be asymmetric when listener A uses listener B's HRTF and vice versa.} 
Future work includes \nz{i) conducting listening experiments to validate improvements noticeable to end users, and ii) extending the proposed loss functions to binaural reproduction~\cite{rafaely2025loss}.}




\section{Conclusions}

We investigated how latent HRTF representations relate to objective perceptual metrics—localization, coloration, and externalization—and introduced an INR‑based framework that embeds HRTFs in a perception-informed space by combining metric MDS supervision with an optional perceptual loss. The proposed method significantly increases the correlation between latent distances and perceptual metrics. HRTF personalization experiments further suggest that our proposed representations improve the selection of subject‑specific HRTFs, pointing toward perceptually-aligned spatial audio rendering.

\vfill\pagebreak



\clearpage
\bibliographystyle{IEEEtran}
\bibliography{refs25}







\end{document}